\documentclass[twocolumn,eqsecnum,aps]{revtex4}

\newcommand{\figurewidth}{84mm}

\usepackage{graphicx}
\usepackage{dcolumn}
\usepackage{amsmath}

\newcommand{\vol}{\Omega}
\newcommand{\OO}{{\mathcal{O}}}
\newcommand{\PP}{{\mathcal{P}}}

\newcommand{\RR}{{\bf R}}
\newcommand{\RRp}{{\bf R'}}
\newcommand{\PPP}{{\bf P}}
\newcommand{\kk}{\mathbf{k}}
\newcommand{\rr}{{\bf r}}
\newcommand{\pp}{\mathbf{p}}

\newcommand{\dd}{{\bf d}}
\newcommand{\sss}{{\bf s}}
\newcommand{\HH}{{\mathcal{H}}}

\newcommand{\ZZ}{\mathcal{Z}}

\newcommand{\beq}{\begin{equation}}
\newcommand{\eeq}{\end{equation}}
\newcommand{\bea}{\begin{eqnarray}}
\newcommand{\eea}{\end{eqnarray}}

\begin{document}

\title[Short Title]{
{\sf \vspace*{-12mm}\normalsize Submitted to Physical Review B, October 16, 2003.}\\[3mm]
Path Integral Monte Carlo Calculation of the Momentum Distribution of
the Homogeneous Electron Gas at Finite Temperature}

\author{B. Militzer}
\email{militzer@gl.ciw.edu}
\affiliation{Geophysical Laboratory, Carnegie Institution of Washington,
      5251 Broad Branch Road,~NW, Washington, DC 20015}

\author{E. L. Pollock}
\email{pollock1@llnl.gov}
\affiliation{Lawrence Livermore National Laboratory,
         University of California, Livermore, CA 94550}

\author{D. M. Ceperley}
\email{ceperley@uiuc.edu}
\affiliation{Department of Physics,
          National Center for Supercomputing Applications,
          University of Illinois at Urbana-Champaign, Urbana, IL 61801}


\begin{abstract}
Path integral Monte Carlo (PIMC) simulations are used to calculate the
momentum distribution of the homogeneous electron gas at finite
temperature. This is done by calculating the off-diagonal elements of
the real-space density matrix, represented in PIMC by open paths. It
is demonstrated how the restricted path integral Monte Carlo methods
can be extended in order to deal with open paths in fermionic systems
where a sign problem is present. The computed momentum distribution
shows significant deviations for strong correlation from free fermion
results but agrees with predictions from variational methods.

\end{abstract}

\maketitle

\section{Introduction}

The momentum distribution is one of the fundamental properties of
a quantum system. It can be directly measured by inelastic
scattering or by studying the trajectories of particles. The first
method has been used to determine the condensate fraction in
liquid helium while the latter was used to demonstrate that
Bose-Einstein condensation of supercooled alkali atoms in magnetic
traps had been achieved.

In this article, we calculate the momentum distribution of fermion
systems at finite temperature with path integral Monte Carlo
(PIMC). The results show how the fermionic momentum distribution
evolves as a function of temperature. At a temperature much higher than
the Fermi temperature, $T_F$, the fluid has a Maxwell-Boltzmann
momentum distribution and with decreasing temperature, the Pauli
exclusion principle becomes more important giving rise to a Fermi
surface.  While it is very simple to study this cross-over in
noninteracting systems, interparticle interactions require a much more
sophisticated description.

Here we compute the effect of interactions of the momentum
distribution at finite temperature using fermion restricted path
integral Monte Carlo simulations. To determine the needed
off-diagonal density matrix elements we need to perform
simulations with open paths. We compute these properties for the
homogeneous electron gas and demonstrate how increased correlation
effects alter the momentum distribution at different temperatures.

\section{Path integral Monte Carlo}

We now give a brief review of the PIMC simulation technique for
fermions and of the restricted path method. In PIMC calculations
of the diagonal density matrix each particle is represented by a
{\em closed path} in imaginary time. Following the procedure
developed for bosonic systems~\cite{Ce95}, we extend fermionic
PIMC to estimate off-diagonal density matrix elements needed for
the momentum distribution. This requires one of the particle paths
to be {\em open}. Special emphasis will be placed on how the path
restriction is applied to open paths and how the Monte Carlo
sampling is affected.

\subsection{Restricted paths technique}

The thermodynamic properties of a quantum many-body system can be
derived from the thermal density, $\hat{\rho} = e^{-\beta
\hat{\HH}}$ with $\beta = 1/k_B T$.  For the purpose of 
performing Monte Carlo simulations, we express this operator in
position-space,
\beq
\rho(\RR,\RR';\beta) \equiv \left< \RR \left| \hat{\rho} \right| \RR' \right>
= \sum_s e^{-\beta \epsilon_s} \, \Psi^*_s(\RR) \, \Psi_s(\RR') \;,
\eeq
where $\Psi_s$ are the many-body eigenfunctions and $\epsilon_s$
the corresponding eigenvalues. $\RR = \left\{ \rr_1, \dots, \rr_N
\right\}$ represents a set of coordinates of $N$ particles in $d$
dimensions. The density matrix of a bosonic (B) or fermionic (F)
system, $\rho_{\rm B/F}(\RR,\RR';\beta)$, can be constructed from
the density matrix for distinguishable particles by a sum of
permutations, $\PP$, to project out states of the corresponding
symmetry, 
\beq 
\rho_{\rm B/F}(\RR,\RR';\beta) = \frac{1}{N!}
\sum_\PP \; (\pm 1)^\PP \; \rho_{\rm D}(\RR,\PP \RR' ; \beta) \;,
\label{BF_rho} 
\eeq
where $(\pm 1)^\PP$ denotes the sign of the permutation. Using the
operator identity, $e^{-\beta \hat{\HH}} = ( e^{-\tau
\hat{\HH}})^M$, the density matrix at temperature $T$ can be
expressed in terms of density matrices at a higher temperature
$MT$. This leads to a path integral with $M$ steps in imaginary
time of size $\tau = \beta / M$,
\bea
\nonumber \label{BF_path_integral}
\rho_{\rm B/F}(\RR,\RR';\beta) \!\!&=&\!\!
\frac{1}{N!} \sum_\PP \; (\pm 1)^\PP \!\!
\int \!\! \ldots \!\! \int \dd\RR_{1}\,\dd\RR_{2} \ldots \dd\RR_{M-1}\;\\
&&
\nonumber
\!\!\!\! \!\!\!\! \!\!\!\! \!\!\!\! \!\!\!\! \!\!\!\! \!\!\!\! \!\!\!\!
\rho_{\rm D}(\RR,\RR_{1}; \tau ) \: \rho_{\rm D}(\RR_{1},\RR_{2};\tau ) \ldots
\rho_{\rm D}(\RR_{M-1},\PP \RR' ;\tau )\\
\!\! &=&\!\!  \frac{1}{N!} \sum_\PP \; (\pm 1)^\PP \! \! \! \! \! \int
\limits_{\RR \rightarrow \PP \RR' } \!\!\!\! \dd\RR_t \;\;
e^{-S[\RR_t] }\;, \!\!\!\! \!\!\!\!
\eea
where $S$ represents the ``action'' of the path $\RR_t$ beginning at
$\RR$ and ending at $\PP \RR'$. Here we use the pair density matrix
for the link actions: $\rho_{\rm D}(\RR_{1},\RR_{2};\tau )$. For
bosonic many-body systems, the integrand is nonnegative and this
expression can be efficiently evaluated using Monte Carlo
techniques~\cite{Ce95}. In the case of fermions, a straightforward
evaluation of this expression is impractical because the cancellation
of many positive and negative terms of the same order leading to
numerically inefficient computation in the low temperature,
many-fermion limit. While one can still use this expression to
numerically study systems of a few fermions, the efficiency
rapidly decays with increasing number of particles and decreasing
temperature, $\sim e^{-\beta N}$. This is referred to as the {\em
fermion sign problem.}

Ceperley~\cite{Ce91,Ce92,Ce96} has shown that the fermion sign
problem in imaginary time path integrals can be solved by {\em
restricting} the path integration to a subvolume of the entire
path space. Let us define the {\em nodes} of the fermion many-body
density matrix as the surface where $\rho_F(\RR,\RR';t)=0$. The
nodes are used to confine the paths $\RR(t)$ to regions where the
density matrix is nonzero, $\rho_F(\RR^*,\RR(t);t) \neq 0$.
$\RR^*$ is called the {\em reference point} and defines the region
in $\{ \RR,t \}$ space ($\Upsilon(\RR^*;t)$) where the path is
allowed to be.  The fermionic density matrix is then given by the
restricted path integral, \beq \rho_F(\RR^*, \RR' ;\beta) =
\frac{1}{N!}\; \sum_\PP \; (-1)^\PP \! \! \! \! \! \! \! \! \! \!
\! \! \! \! \! \! \! \! \! \! \! \! \! \! \!
\int\limits_{\quad\quad\quad\quad \RR^* \rightarrow \PP \RR' \in
\Upsilon(\RR^*;t)} \! \! \! \! \! \! \! \! \! \! \! \! \! \! \! \!
\! \! \! \! \! \! \! \! \! \! \dd\RR_t \;\; e^{-S[\RR_t] } \quad,
\label{restricted_PI} \eeq
where one sums and integrates over all paths that never cross the nodal
boundaries. By introducing this restriction, one effectively cancels
all negative and some positive contributions to the trace of density
matrix. Complete cancellation of all negative terms, however, is only
reached on the diagonal of the density matrix.  Off the diagonal,
negative contributions also enter the restricted path integral, as
will be illustrated below. In either case, the restriction gives rise
to an efficient numerical algorithm that scales favorably with
increasing number of particles, similar to that for bosons.

The expression in Eq.~\ref{restricted_PI} is exact as long as the
restriction is exact~\cite{Ce91}. The exact density matrix is
only known in a few cases, e.g. for noninteracting particles. In
practice, one introduces a {\em trial} density matrix $\rho_{\rm
T}(\RR,\RR';\beta)$ that provides approximate fermion nodes which
introduces an approximation in computed observables.  However, for
many systems this technique has worked well.

The simplest approximation for the trial density matrix
$\rho_T(\RR,\RR';\beta)$ is a Slater determinant of single
particle density matrices,
\beq
\rho_T(\RR,\RRp;\beta)=\left|
\begin{array}{ccc}
\rho^{[1]}(\rr_{1},\rr'_{1};\beta)&\ldots&\rho^{[1]}(\rr_{N},\rr'_{1};\beta)\\
\ldots&\ldots&\ldots\\
\rho^{[1]}(\rr_{1},\rr'_{N};\beta)&\ldots&\rho^{[1]}(\rr_{N},\rr'_{N};\beta)
\end{array}\right|
\quad. \label{matrixansatz}
\eeq
For a homogeneous Fermi liquid,
such as the electron gas, the single particle density matrices
are:
\beq
\label{single_gaussian}
\rho^{[1]}_0(\rr,\rr';\beta) = (4 \pi \lambda
\beta)^{-d/2} \exp \left \{ -\frac{(\rr-\rr')^2}{4\lambda \beta}
\right \}
\quad,
\eeq
where $\lambda=\hbar^2/2m$. For temperatures above the Fermi
energy, where exchange is small, the nodal surfaces of this
determinant are accurate. At decreasing temperature, interaction
effects become more important and consequently the error from
employing free particle nodes increases. (Note, however, that the
nodes are second order in the interaction.) One can improve the
nodal approximation by using the dual-reference point
method~\cite{Ce96}. In this approach, the nodal restriction is
$\rho_T(\RR(t),\RR^*; t^*)\neq 0$ where $t^*=\min(t,\beta-t)$ is
the smaller imaginary time separation from the reference point.
The dual-reference point method will be used throughout this work.

How sensitive the derived results are to the accuracy of the nodes
can depend on the interactions in a particular system. Generally,
one expects free particle nodes to work well at high temperature
and when correlation effects are weak. Also, when particles are
localized like the electrons in molecular hydrogen or particles in
a Wigner crystal, the effect of the nodes is reduced and nodal
restriction is unimportant. One can improve nodal surfaces in
several ways. For example, one can derive the nodes from a
variational density matrix~\cite{MP00}, by treating the
interactions with Hartree-Fock. Or, one can use backflow in the
density matrix to account for interaction
effects~\cite{Ce91,Ce96}.


Consider the calculation of an observable, $\OO$, diagonal in $\RR$
space such as the pressure, kinetic, potential and internal energy
as well as pair correlation functions:
\bea
\label{expectation_value} \left< \OO \right> &=& \frac{1}{\ZZ}
\int \! \dd\RR  \; \rho(\RR,\RR;\beta) \; \left< \RR \right| \OO \left| \RR \right>\;,\\
\ZZ &=& \;\;\;\; \int \! \dd\RR \; \rho(\RR,\RR;\beta) \quad. 
\eea
Such computations require only simulations with closed paths,
beginning at a point $\RR$ and ending at its permutation, $\PP
\RR$. The restriction eliminates all contributions from odd
permutations that would enter with a negative weight because their
paths would violate the nodal constraint an odd number of times.

\subsection{Computation of the momentum distribution}

The single-particle momentum distribution for $N_\sigma$ particles in
spin state $\sigma$ is defined as,
\beq
n(\kk) = \frac{(2\pi \hbar)^d}{\vol} 
        \left< \sum_{j=1}^{N_\sigma} \delta( \hat{\pp}_j-\hbar \kk) \right>\;,
\eeq
with the normalization for a finite system and in the
thermodynamic limit respectively given by,
\beq 
\sum_{\kk} n(\kk) = N_\sigma 
\;\;\;\;,\;\;\;\;
\frac{\vol}{(2\pi)^d} \int \dd\kk \; n(\kk) = N_\sigma  \;.
\eeq 
Inserting complete sets of states and using $\left< \RR | \PPP \right>
=e^{-i \RR \cdot \PPP / \hbar} / (2\pi\hbar)^{Nd/2}$, one finds,
\bea
n(\kk)\!\!&=&
\nonumber \!\!
         \frac{1}{\ZZ \vol} \int \!\! \dd\RR \dd\RR' \dd\PPP
         \left<\RR|\hat{\rho}|\RR'\right>
         \frac{e^{i\PPP\cdot(\RR-\RR')/\hbar}}{(2\pi\hbar)^{(N-1)d}}
         \sum_{j=1}^{N_\sigma} \delta(\pp_j-\hbar\kk)\\
\nonumber && \!\!\!\!\!\!\!\!\!\!\!\!\!\!\!\!\!\!\!\!\!\! = 
         \frac{N_\sigma}{\ZZ\vol} \int \!\! \dd\RR \dd\rr'_{1}
         e^{i(\rr_1-\rr'_1) \cdot \kk}
         \rho(\rr_1, \ldots, \rr_N,\rr'_1,\rr_2,\ldots,\rr_N).
\eea 
The last line follows from the equivalence of all the particles
assuming particle 1 has spin $\sigma$. Consequently, $n(\kk)$ is given
by the Fourier transform of the single-particle reduced density matrix
(SPRDM), $n(\sss)$,
\bea
n(\kk) &=& \frac{N_\sigma}{\vol} \int \dd\sss \; e^{-i \kk \sss} \; n(\sss)
\;\;, \\
%
\nonumber
n(\sss) &=& \frac{1}{\ZZ} \int \dd \RR \:
\rho(\rr_1,\rr_2,\ldots,\rr_N \, , \,\rr_1+\sss,\rr_2,\ldots,\rr_N) \;. \label{rho1}
\eea
Note that $n(s=0)=1$ from the definition. 

Classical particles have a Maxwellian momentum distribution with
\bea
n(\kk) &=& \frac{N_\sigma}{\vol}  (4 \pi \lambda \beta)^{d/2} \exp
\left\{-\beta \lambda \kk^2\right\}\;,
\label{Gauss_nk}\\
n(\sss) &=& \exp \left\{ -\frac{\sss^2}{4 \lambda \beta }\right\}
\label{Gauss_nr}
\;.
\eea
For an ideal Fermi gas in 3 dimensions at $T=0$, the momentum
distribution (for one spin state) is a Fermi function,
\bea
\nonumber
n(\kk) &=& \left\{
\begin{array}{cl}
 1             & ~~{\rm{for}}~~~ k\leq k_F ~~~~~{\rm{with}}~~k_F = (6 \pi^2 N_\sigma/\vol )^{1/3}\\
 0             & ~~{\rm{for}}~~~ k > k_F\quad
\label{nofk_ideal}
\end{array}\right.\\
\nonumber 
\label{nofr_ideal} 
n(\sss) &=& 3/x^3 
  \left[ \sin \, x - x \, \cos \, x \right] 
  ~{\rm{with}}~~ x = s \, k_F \quad. 
\eea
Hence, the free fermion SPRDM decays to zero as $\cos(s \,
k_F)/s^2$ at zero temperature.

The function $n(\sss)$ can be computed from PIMC simulations with
one open path, where the vector $\sss$ is the separation of the
two open ends. Methods developed for bosonic systems~\cite{Ce95},
such as liquid $^4$He, carry over directly to fermion path
integrals except for the considerations having to do with the
nodal restrictions. In the simulations, $n(s)$ is obtained as a
histogram, in which one enters the observed separations of the
open ends weighted with the sign of the permutation. The
discussion in the previous subsection concerning the elimination of
odd permutations holds for diagonal paths. Off-diagonal paths with
odd permutations do contribute to the Monte Carlo averages with a
negative weight. At separations where $n(s)$ is negative, odd
permutations outweigh even permutations. The algebraic decay of
$n(\sss)$ requires long exchange cycles, of the order of the
number of particles. In restricted PIMC, there is a direct
relation~\cite{Ce96} between long exchange cycles and the
discontinuity of $n(k)$ at $k=k_F$.

\subsection{Example of two free fermions}

\begin{figure}[htb]
\includegraphics[angle=0,width=\figurewidth]{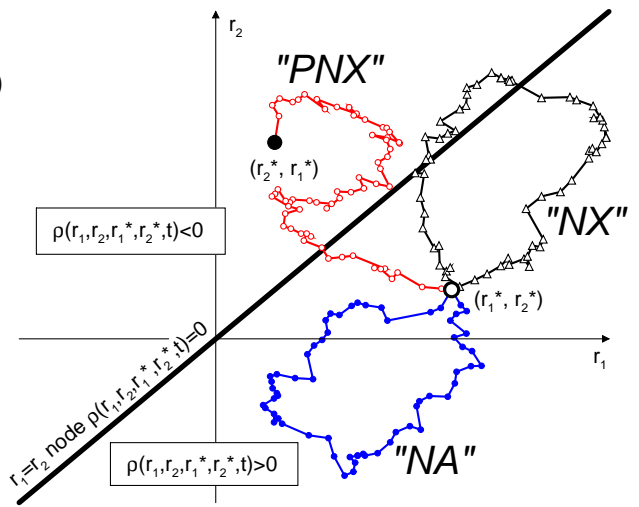}
\caption{Illustration of the three different types of paths for
         two free fermions all starting
         at a given reference point $(\rr_1^*,\rr_2^*)$. A
         node-avoiding (``NA''), a node-crossing (``NX'') as well as
         a permuting and consequently node-crossing path (``PNX'') are
         shown in the $(\rr_1,\rr_2)$ plane. The thick solid line
         indicates the node.}
\label{closed_paths}
\end{figure}

\begin{figure}[htb]
\includegraphics[angle=0,width=\figurewidth]{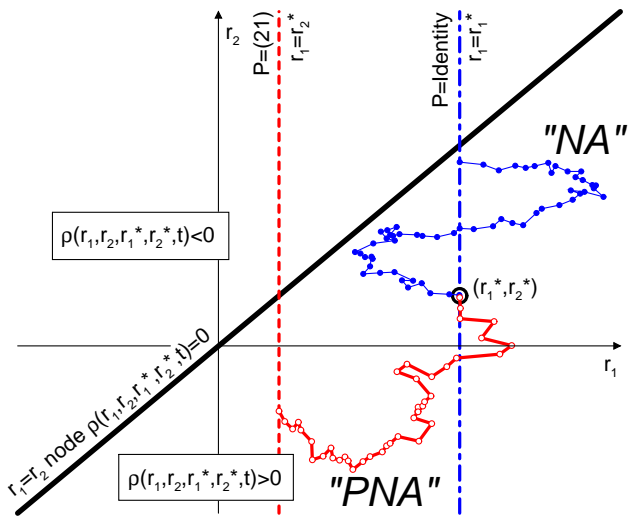}
\caption{Illustration of node-avoiding paths that contribute to
         {\em off-diagonal} density matrix elements for two free
         fermions. All start at the reference point
         $(\rr_1^*,\rr_2^*)$. Shown are paths with one open end at
         $r_2$. The dot-dashed line indicates possible end points for
         nonpermuting paths [$\rr_1=\rr_1^*$, labeled ``NA'']. In
         contrast to {on-diagonal} matrix elements shown in
         Fig.~\ref{closed_paths}, here, we find permuting but
         node-avoiding paths with end points on the dashed line
         [$\rr_1=\rr_2^*$, labeled ``PNA''] that represent negative
         contributions to the density matrix.}
\label{open_paths}
\end{figure}

In order to illustrate the restricted path integral technique, we
discuss the simplest fermionic system: the case of two free
fermions with the same spin without boundaries.
The density matrix $\rho_{\rm T}(\rr_1,\rr_2,\rr_1^*,\rr_2^*;\beta)$
(Eq.~\ref{matrixansatz}) is positive when
\beq
     (\rr_1 - \rr_2) \cdot (\rr_1^* - \rr_2^*) > 0.
\eeq 
The nodal surface is the hyperplane $\rr_1=\rr_2$ at all
temperatures as illustrated in Fig.~\ref{closed_paths}. This
figure also shows the three types of paths that contribute to the
trace of the density matrix. Since we only consider diagonal
density matrix elements in this figure, the path starting at the
reference point, $(\rr_1^*,\rr_2^*)$, can either return to its
origin or to the only possible permutation of it,
$(\rr_2^*,\rr_1^*)$.

In restricted PIMC, only node-avoiding paths, labeled ``NA'',
contribute. In the case of two particles, the restriction prohibits
any permutation. In case of the direct (unrestricted) fermion method,
nonpermuting path that cross the nodes (``NX'') and those that avoid
it, enter with a positive sign. Permuting paths (``PNX'') now also
enter with a negative weight because of the $(-1)^\PP$ factor. For
bosons, permuting and nonpermuting paths enter a with positive weight.

In order to compute the momentum distribution for two particles,
one would perform simulations with one open and one closed path.
Fig.~\ref{open_paths} illustrates such paths in the
$(\rr_1,\rr_2)$ plane. For simplicity, only node-avoiding paths
that begin at $(\rr_1^*,\rr_2^*)$ and have an open end at
$\rr_2$ are shown. Under these restrictions, only two types
of paths remain: ($i$) Nonpermuting paths with
$\rr_1=\rr_1^*$ and $\rr_2 \ne \rr_2^*$. Path 1
would appear as a closed polymer while path 2 would be open.
($ii$) A new category appears: node-avoiding but permuting paths
with $\rr_1=\rr_2^*$ and $\rr_2 \ne \rr_1^*$. In
the case of closed paths, the nodal restriction eliminates
permutations because the final point $(\rr_2^*,\rr_1^*)$ lies on
the other side of the node. However, for open paths, many final
points for permuting paths are possible (see dashed line in
Fig.~\ref{open_paths}) because path 2 does not have to end at the
beginning of path 1.

With increasing number of fermions, it is difficult to illustrate
the nodal constraint graphically. However, this is not a
computational difficulty since one can easily verify the sign of
the Slater determinant for any proposed path to see if it is
node-avoiding.

\subsection{Monte Carlo sampling with open paths}

To apply the restricted path integral to open paths, one first has to
decide at which time slice one opens the paths, relative to the
reference point, assumed to be at $t=0$.  Using the dual reference
point method, one can put the open ends at $t=0$, or at
$t=\beta/2$. We used the latter choice because it is more symmetrical
and it avoids the singular behavior of the fermion density matrices as
$t\rightarrow 0^+$.

To move the open and closed paths in the presence of nodal
constraints, one first picks a time interval of size $2^l$ in which
the paths will be modified where $l$ is the number of levels in the
bisection method~\cite{Ce95}. If it is chosen too large, most trial
moves will be rejected. If it is too small, the paths diffuse too
slowly. The optimal choice depends on strength of the interactions and
on the fermion degeneracy. After selecting the time interval, one uses
the heat-bath algorithm (see~\cite{Ce95} for details) to sample a
particular cyclic permutation of moving particles. Then one proceeds
with a bisection method for all $l$ levels in order to determine the
positions of the moving particles at each time slice. In this study,
the sampling probabilities are taken from the free particle density
matrix.
For a closed path connecting points $\rr_1$ and $\rr_2$, this becomes,
\beq
T_l(\rr) = ( 2^l \pi \lambda \tau )^{d/2} \exp \left
\{ -\frac{ (\rr-\rr_{\rm m})^2}{2^l \lambda \tau} \right \} \;\;,
\eeq
where $\rr_{\rm m}$ is the mid point $(\rr_1+\rr_2)/2$. For strongly
interacting systems, a better sampling efficiency has been reached by
adding a drift and a covariance term~\cite{Ce95}. For an open end,
which is only connected to one point $\rr_1$, the free particle
sampling probability is,
\beq 
T_l(\rr) = ( 4^l \pi \lambda \tau )^{d/2}
\exp \left \{ -\frac{ (\rr-\rr_1)^2}{4^l \lambda \tau} \right \} \;\;.
\eeq
The pair action, $ u(\RR,\RR';\beta)=-\log[\rho(\RR,\RR';\tau)/
\rho_0(\RR,\RR';\tau)]$, is then used to determine whether a
particular configuration will be rejected or temporarily accepted.

As a final step, one verifies if the nodal constraint is satisfied
at each time slice. Otherwise, the proposed configuration is
rejected. For fermion simulations with only closed paths, one can
eliminate odd permutation moves because they would inevitably
violate the nodes. In the presence of open paths, this is not
true; otherwise we would not get the negative pieces of the
density matrix. The node-avoiding condition for each time slice is
still the same, $\rho_T(\RR(t),\RR^*;\min(t,\beta-t))>0.$  The
following two conditions can be used to eliminate paths violating
the nodal constraint: (a) It is impossible to permute an even
number of closed paths while keeping all other particle
coordinates fixed. (b) It is impossible to permute an open and a
closed path in a move that does not change the time slice with the
open ends. It should be noted that the rules are only employed to
improve the efficiency. Paths violating these conditions would be
rejected anyway when the nodal constraints are enforced.

The SPRDM is proportional to distribution of $\sss$, the separation of
the two ends.  Configurations with an odd permutation contribute
negatively. For a finite homogeneous system, $n(\sss)$ is only a
function of $|\sss|$. However, in a periodic but finite simulation
cell, there is a dependence on the orientation of $\sss$ with respect
to the cell boundaries. In all following results, $n(\sss)$ will be
spherically averaged. However, for the computation of the momentum
distribution, $n(\kk) \propto \left < e^{i\kk \sss} \right>$, the
angular dependence of $n(\sss)$ is important. We compute $n(\kk)$
directly during the Monte Carlo simulation for a reasonable number of
$\kk$ vectors, to avoid storing the $d$ dimensional function
$n(\sss)$.

The normalization of $n(s)$, is not determined {\em a priori} with
a single PIMC run. We use the same method developed for the
bosonic superfluid helium. The normalization is determined by its
value at $n(s=0)=1$. A difficulty arises from the fact that the
probability that the two ends have a distance less than $s$ scales
as $s^d$ in $d$ dimensions and error bars accordingly increase for
small $s$ as $s^{-d/2}$.  To smooth the errors, we fit the
observed histogram of end-to-end separations to a function of the
form:
\beq
\lim_{s \to 0} n_{\rm MC} (s) = \xi_1 \left[ 1 - \frac{\left<K\right>}{\lambda d}
s^2 + \xi_2 s^4 \right]\;\;,
\eeq
for the unknown parameters $\xi_1$ and $\xi_2$. Here $K$ is the kinetic
energy derived from a separate simulation with only closed paths.
The specific values are given in Table ~\ref{tab1} together with
potential energy, $V$. The internal energy can be obtained by
adding $K+V$ and pressure for coulomb systems, follows from the
virial theorem, $3P\vol = 2K+V$. The fit then determines the
normalization constant $\xi_1$ of the momentum distribution. To
further reduce the error bars of $n(s)$, we enhanced the frequency
of moving the open ends by selecting the open path more often and
also by moving the slice with the open ends more frequently.

\begin{table}
\caption{Kinetic and potential energy, $K$ and $V$, per particle in units of Hartrees are tabulated
    for  $r_s=4$ and 40 for the spin polarized
    ($N_\uparrow=0$) and unpolarized homogeneous electron
    gas ($N_\uparrow=N_\downarrow$). }
\begin{tabular}{ccc|c|c|c|c}
$\frac{T_F}{T}$ & $N_\uparrow$ & $N_\downarrow$ & $K[r_s=4]$ & $V[r_s=4]$ & $K[r_s=40]$ & $V[r_s=40]$ \\
\tableline
1  & 33 & 0 &  0.2961(6) &  --0.1513(1)  & 0.00354(2)~~& --0.019553(4) \\
2  & 33 & 0 &  0.177(1)~~&  --0.16413(6) & 0.002605(6) & --0.019774(2) \\
4  & 33 & 0 &  0.144(4)~~&  --0.16997(5) & 0.00237(1)~~& --0.019826(2) \\
8  & 33 & 0 &  0.143(8)~~&  --0.17223(6) & 0.00232(3)~~& --0.019834(2) \\
16 & 33 & 0 &  0.14(1)~~~&  --0.17341(4) & 0.00227(4)~~& --0.019834(4) \\
\tableline
1  & 33 & 33 & 0.1930(5)  & --0.15574(9)  &  0.002859(4)   & --0.019715(2) \\
2  & 33 & 33 & 0.123(1)~~ & --0.1627(1)~~ &  0.002389(5)   & --0.019789(3) \\
4  & 33 & 33 & 0.105(3)~~ & --0.1636(1)~~ &  0.00230(1)~~  & --0.019803(2) \\
8  & 33 & 33 & 0.101(6)~~ & --0.1632(2)~~ &  0.00227(3)~~  & --0.019803(2) \\
16 & 33 & 33 & 0.100(8)~~ & --0.1631(2)~~ &  0.00232(5)~~  & --0.019797(6) \\
\end{tabular}
\label{tab1}
\end{table}

\section{Results}

\begin{figure}[htb]
\includegraphics[angle=0,width=\figurewidth]{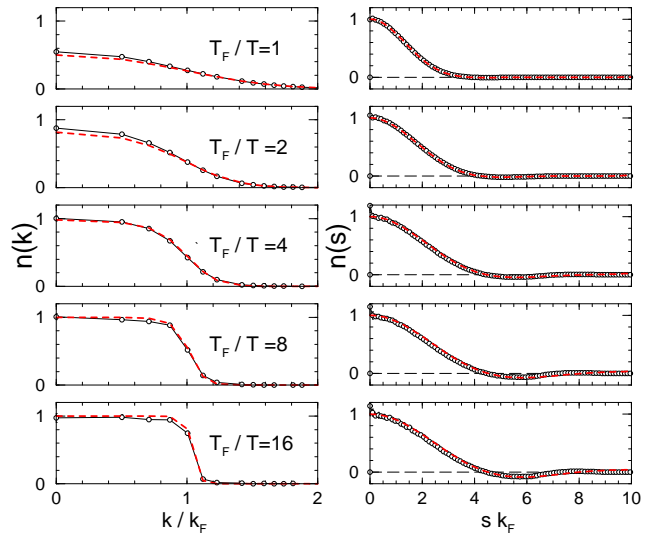}
\caption{The left panels show the momentum distributions $n(k)$
       from PIMC simulations (circles) with $33$ spin polarized
       electrons for different temperatures at ($r_s=4$). For $T \geq
       T_F/4$, the population of low momentum states is enhanced
       compared with the corresponding ideal Fermion results (dashed
       lines), which leads to a lowering of kinetic energy
       ($K<K_0$). The right panel shows the corresponding off-diagonal
       density matrix elements $n(r)$. With decreasing temperature, a
       negative region near $s k_F=5.8$ is apparent.}
\label{mom_rs4_pol}
\end{figure}

\begin{figure}[htb]
\includegraphics[angle=0,width=\figurewidth]{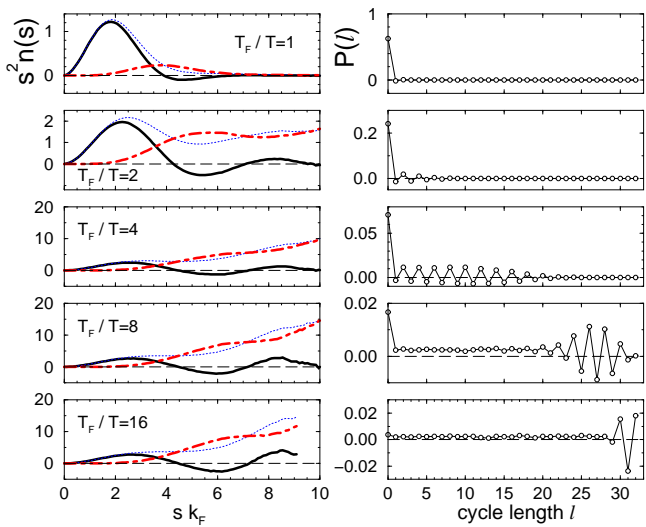}
\caption{ The left column shows the distribution of the open ends,
        $n(s)$ multiplied by the volume element $s^2$ (solid lines)
        for a system of 33 spin polarized electrons for $r_s=4$ at
        different temperatures (see corresponding
        Fig.~\ref{mom_rs4_pol}). The dash-dotted lines indicate the
        distribution of paths that enter with a negative sign. Their
        contribution vanishes more quickly for small $s$ because they
        do not contribute to diagonal density matrix elements. The
        dashed lines indicate the distribution of positive
        contributions. The right column shows the distribution of
        permutation cycles as a function of cycle length $l$ times the weight of the
        overall permutation. }
\label{cycles}
\end{figure}

\begin{figure}[htb]
\includegraphics[angle=0,width=\figurewidth]{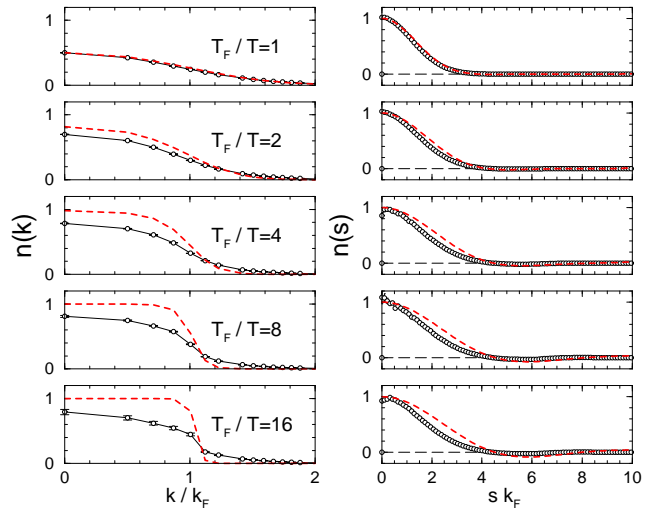}
\caption { Momentum distribution $n(k)$ and off-diagonal density
         matrix $n(r)$ for the spin polarized electron gas at $r_s=40$
         from simulations with 33 particles. Deviations from the ideal
         Fermion results are increased compared to
         Fig.~\ref{mom_rs4_pol} (see description and line styles
         there) due to increased correlation effects.}
\label{mom_rs40_pol}
\end{figure}

\begin{figure}[htb]
\includegraphics[angle=0,width=\figurewidth]{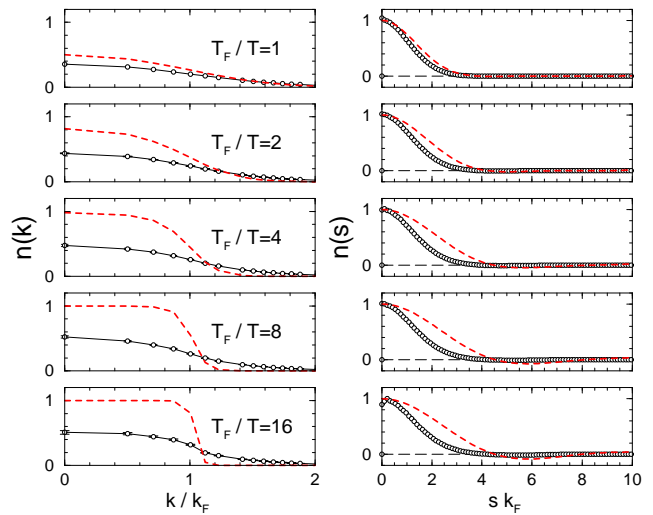}
\caption{Momentum distribution $n(k)$ and off-diagonal density
       matrix $n(r)$ function for the unpolarized electron gas at
       $r_s=40$ from simulations with 66 particles. Correlation
       effects are enhanced compared to Figs.~\ref{mom_rs4_pol} and
       \ref{mom_rs40_pol} (see description and line styles there).}
\label{mom_rs40_nonpol}
\end{figure}

\begin{figure}[htb]
\includegraphics[angle=0,width=\figurewidth]{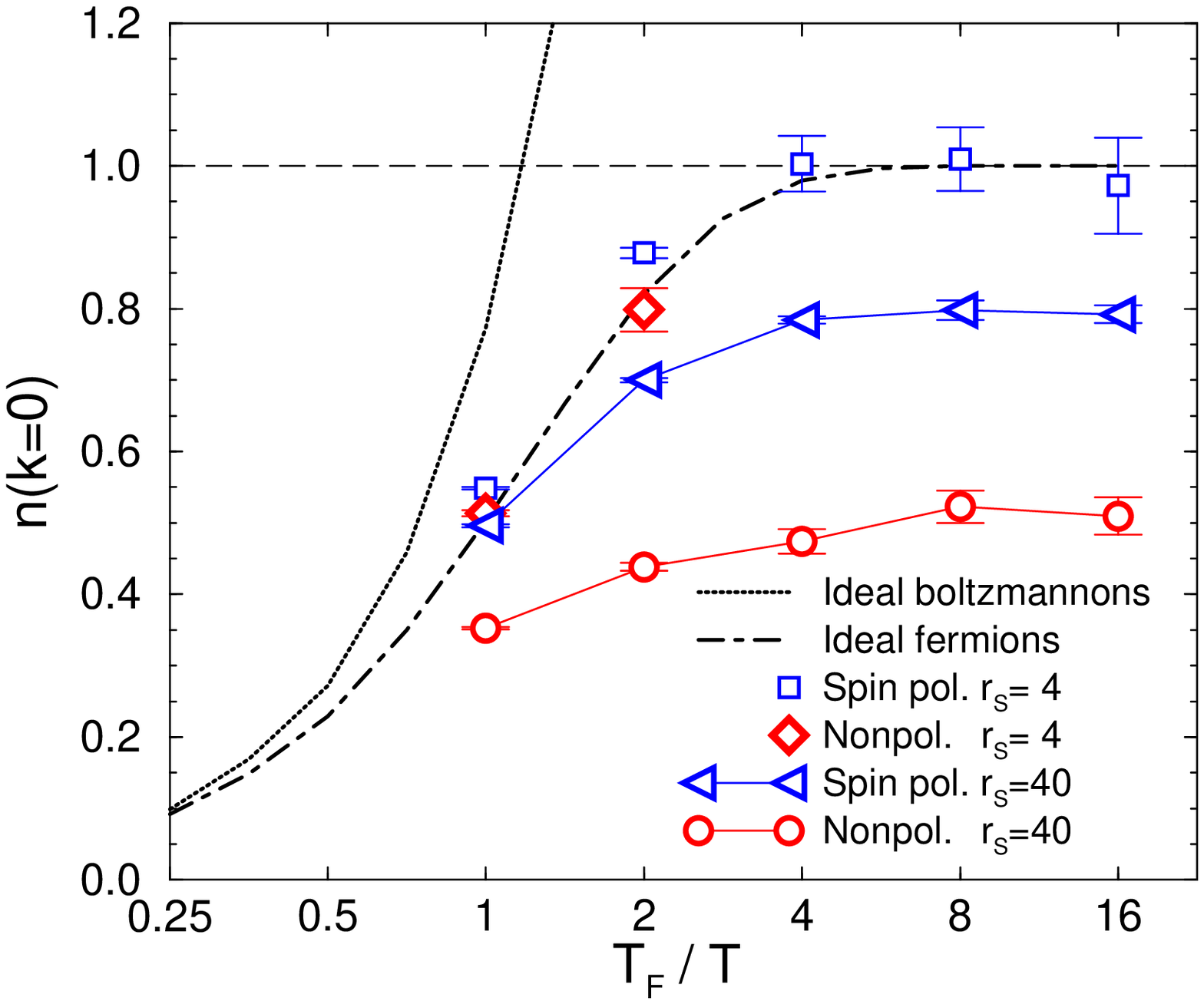}
\caption{The temperature dependence of the population of the
         zero momentum state, $n(k=0)$, is shown for two densities
         ($r_s=4$ and 40) of the fully spin polarized electron gas
         (from PIMC simulations with 33 particles) and for the
         nonpolarized case (simulations with 66 particles).}
\label{n_k_0}
\end{figure}

We determined the momentum distribution of the homogeneous electron
gas at three different conditions. First, we study the spin polarized
electron gas at a density of $r_s=4$, which corresponds to the
electron density of a low density metal such as sodium. The computed
momentum distribution turns out to be similar to an ideal Fermi
gas. Secondly, we calculated $n(k)$ for spin polarized electron gas at
a much lower density of $r_s=40$, where correlation effects are very
large and significant deviations from free particle behavior are
found. Finally, we present results for the unpolarized electron gas at
$r_s=40$ where correlation effects are even more important. For all
three conditions, we have computed the momentum distribution for a
series of different temperatures ranging from $1 \leq T_F/T \leq
16$. We compare with results from zero temperature quantum Monte Carlo
simulations.

Fig.~\ref{mom_rs4_pol} shows the momentum distribution for spin
polarized electron gas as a function of temperature. In the limit of
high temperature $T \gg T_F$, fermion effects become less important
and one recovers the classical Maxwell-Boltzmann distribution. With
decreasing temperature, one finds that population of low momentum
states with $k<k_F$ increases until it reaches 1, the maximum allowed
by the Pauli exclusion principle.  Simultaneously, the slope at the
Fermi wave vector becomes increasingly steep. We do not exactly
recover the limit of ideal Fermi function in Eq.~\ref{nofk_ideal},
because of finite size effects. The system size we used, $N=33$ at
$T=T_F/16$, is already a demanding computation, in part because we
used a time step of $\tau=1/32 T_F$ needed to enforce the nodal
constraint accurately along the paths but requiring simulations with
up to 384 time slices. Notice that at $T=T_F/16$ there are still small
but nonnegligible thermal excitations of states above $k_F$ present.
The correlation effects, absent for free particles, are small but
nevertheless significant. At high temperature, the interactions lead
to an increased population of low momentum states resulting in a
lowering of the total kinetic energy. The reason for this effect is
that the entropy is the dominant part of the free energy at high
temperature. Interactions can lower the entropy which also leads to a
lowering of the kinetic energy. This effect has been discussed in
detail in~\cite{MP02}. At low temperature, the free energy is
dominated by the interaction term and the kinetic energy is always
higher than the corresponding ideal value. As a result, even at $T=0$,
states above the $k_F$ are populated. According to Migdal's
theorem~\cite{migdal}, as long as the system remains a Fermi liquid,
the discontinuity at $k_F$ remains, but the step size is reduced
compared to the free fermion value.

We have shown the SPRDM $n(s)$ on the right side of
Fig.~\ref{mom_rs4_pol}; they are very close to those of free
particles at this density. However, in the more correlated
systems, discussed later, significant deviations from the ideal
behavior are found. At high temperature, (in the classical limit)
$n(s)$ is dominated by a single Gaussian. With decreasing
temperature, a shallow negative region develops around $s
k_F=5.8$. At $s k_F=7.25$, $n(s)$ becomes positive again and
exhibits a maximum at $s k_F=8$. Further oscillations cannot be
identified for this system size.

Fig.~\ref{cycles} shows the distribution of positive and negative
contribution of the SPRDM. (The contribution to $s^2 n(s)$ is shown.)
At small separation, $n(s)$ is dominated by the positive contribution
because negative terms are prevented by the restriction since the
paths become diagonal in this limit. The negative region near $s
k_F=5.8$ develops because two particle permutations occur with
increasing probability for temperatures $T \leq T_F$. Such open paths
can spread out further than single open paths and therefore start to
dominate at larger $s$. As the temperature is decreased further below
$T_F$, longer and longer permutation cycles contribute to $n(s)$. The
magnitude of positive and negative contributions increases with $s$
but each function dominates for different separations giving rise to
the oscillatory behavior of the total $n(s)$ function, which is
expected from the zero temperature result.

Fig.~\ref{cycles} also shows the probability of finding a
permutation cycle times the overall sign as a function of the
cycle length. At high temperature the thermal De Broglie wave
length, $\lambda_{\rm th}^2 = 4 \pi \lambda \beta$, is short
compared to the inter-particle spacing, which makes the longer
permutation cycles occur with small probability. With decreasing
temperature, longer permutations occur more frequently and the
cycle distribution approaches a positive constant. The occurrence
of a particular cycle length is no longer correlated with total
permutation, and overall there are more positive than negative
total permutations.  The difference gets smaller as the negative
contributions become more important at an even lower temperature.
This distribution shows oscillation for the largest occurring
cycle lengths since if such a long cycle occurs it is unlikely
there is also another permutation cycle  that can alter the total
sign of the permutation.

Fig.~\ref{mom_rs40_pol} shows the $n(k)$ and $n(s)$ functions for the
spin polarized electron gas at a much lower density of $r_s=40$. Under
these conditions, the particles are significantly more correlated and
their behavior differs substantially from free fermions. As a result,
the momentum distribution is much more spread out leading to the
population of higher momentum states. The total kinetic energy is
clearly above the corresponding ideal value. In the low temperature
limit, the population of the $k=0$ state reaches a value of only about
0.8 relative to free particles. The $n(s)$ distribution is shifted to
smaller $s$ values indicating that the paths are more localized due to
the repulsive interactions.

Fig.~\ref{mom_rs40_nonpol} shows the corresponding results for the
unpolarized electron gas with 66 electrons. These simulations are
comparatively easier than simulations at higher density because the
interactions lead to some localization and cut down on the number of
long permutation cycles which give a fluctuating sign.  Simulations of
a correlated system are therefore less demanding than that of weakly
interacting particles.

\begin{figure}[htb]
\includegraphics[angle=0,width=\figurewidth]{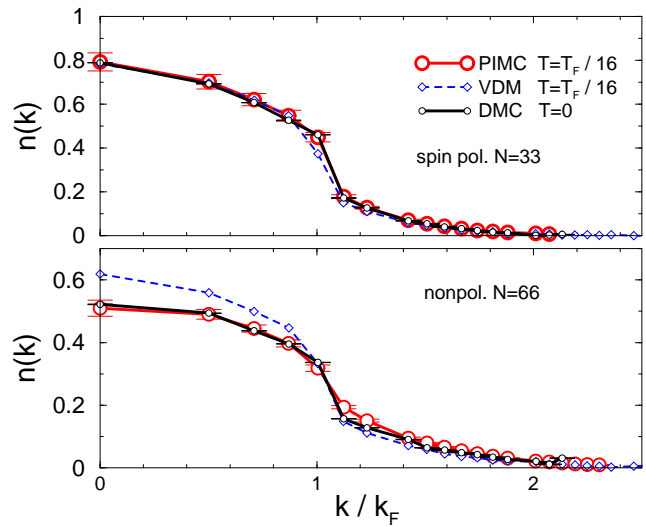}
\caption{Comparison of the momentum distribution for $r_s=40$ at $T=T_F / 16$
     computed with PIMC, VDM and ground state diffusion Monte Carlo (DMC) calculations at
     $T=0$. The upper graph shows results from simulation with
     33 spin polarized electrons, the lower graph represents the
     unpolarized case with 66 particles.}
\label{zeroTlimit}
\end{figure}

\begin{figure}[htb]
\includegraphics[angle=0,width=\figurewidth]{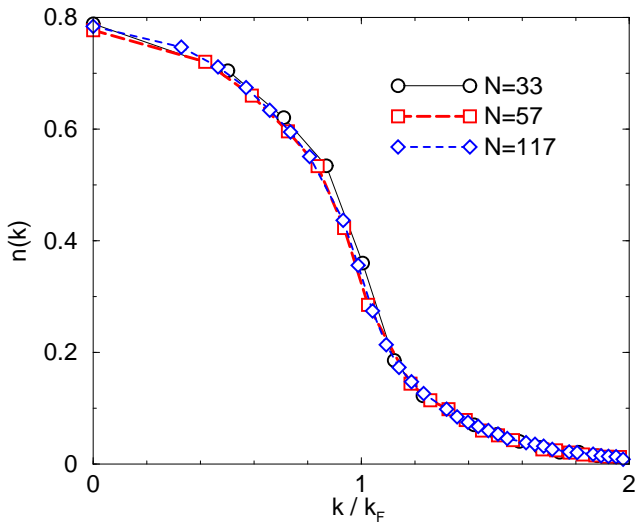}
\caption{Finite size study of the momentum distribution for
    spin polarized electron gas at $r_s=40$ and $T_F/T=8$ computed
    using the variational density matrix method for $N=33$, 57,
    and 117 particles in periodic boundary conditions.}
\label{finite_size}
\end{figure}

In Fig.~\ref{n_k_0}, the occupation of the zero momentum state is
plotted as a function of temperature. At high temperature, all
curves converge to the free particle result. Simulations of the
spin polarized electron gas at $r_s=4$ are above the ideal result
for high temperatures underlining the lowering of the kinetic
energy. For low temperatures, they converge to the ideal value of
1 within the error bars. The graph also shows that the low density
results converge to a ground state limit as well.

In Fig.~\ref{zeroTlimit}, we compare PIMC results and variational
density matrix calculations (VDM)~\cite{VDM} at $T=T_F/16$ with the
ground state momentum distribution derived from diffusion Monte Carlo
(DMC) simulations using backflow nodes~\cite{Zo02} (For additional DMC
results see~\cite{OB94}). The agreement between PIMC and DMC is
excellent for both spin polarizations considered here.  Only for the
nonpolarized case, one observes some very small deviations around $k
\approx k_F$, an indication of thermal excitation present.
Fig.~\ref{zeroTlimit} for the spin polarized case also shows good
agreement with a VDM calculation using the free particle density
matrix (Eq.~\ref{matrixansatz}) multiplied by a temperature dependent
Jastrow factor derived from the random phase
approximation~\cite{VDM}. However, for the unpolarized system, VDM
predicts $n(k)$ values that are significantly too high for
$k<k_F$. While the VDM method can certainly be improved by choosing
parameters in the Jastrow factor~\cite{VDM} more appropriately, it
underlines the need for methods that work without input from
analytical calculations.

To estimate the finite size effects, we used the VDM method which
is significantly less computationally demanding than PIMC.
Fig.~\ref{finite_size} shows the finite size dependence of the
momentum distribution for the spin polarized electron gas at
$r_s=40$ and $T_F/T=8$, a density at which PIMC and VDM agree
well. The simulations were performed for different numbers of
particles corresponding to filled $k$ shell structures with
$N=33$, 57, and 117 particles.  Using periodic boundary
conditions, the function $n(k)$ can only be computed for $k$
values in the reciprocal lattice of the simulation cell.
Consequently, $n(k)$ is shown for a different set of $k$ values
depending on the number of particles. The overall agreement of the
computed momentum distribution is very good, indicating that the
finite size errors are small at these temperatures.
\vspace*{5mm}

\section{Conclusions}

This computational technique allows one to calculate the momentum
distribution within PIMC for fermion systems. It combines the
sampling methods using open paths developed for bosonic liquids
with restricted path technique derived for fermions. Results for
the homogeneous electron gas show that the temperature dependence
of the momentum distribution can be studied and ground state
results can be reproduced. The method is applicable to any Fermi
system, in particular to hot dense hydrogen~\cite{We96} where one
expects significant changes in the momentum distribution with
increasing density as the electrons are delocalized in the
molecular-metallic transition, and to calculate the momentum
distribution of $^3$He~\cite{Se01} and $^3$He-$^4$He mixtures.

\begin{acknowledgments}
The authors acknowledge useful discussions with J. Shumway. This
work was performed in part under the auspices of the U.S.  Dept.
of Energy at the University of California/Lawrence Livermore
National Laboratory under contract no. W-7405-Eng-48 and by NSF
-DMR01-04399.
\end{acknowledgments}


\end{document}